\documentclass{elsart}
\usepackage{graphicx,amsmath,amssymb,bm}
\usepackage{amsfonts,amssymb,amscd,amsmath}
\usepackage{graphicx}

\newcommand{\slashit}[1]{#1 \kern-.45em\slash}

\newcommand{\slashP}{P \kern-.65em\slash }

\begin{document}
\begin{frontmatter}
\title{Azimuthal Asymmetry of Prompt Photons in Nuclear Collisions}
\author{B. Z. Kopeliovich}$^{1,2}$,
\author{A. H. Rezaeian}$^1$ and
\author{Iv\'an Schmidt}$^1$
\address{$^1$ Departamento de F\'\i sica y Centro de Estudios Subat\'omicos, Universidad
T\'ecnica Federico Santa Mar\'\i a, Casilla 110-V, Valpara\'\i so,
Chile \\
$^2$ Joint Institute for Nuclear Research, Dubna, Russia}

\begin{abstract}

\noindent 
The azimuthal elliptic asymmetry v2 observed in heavy ion
collisions, is usually associated with properties of the medium
created in the final state.
We compute the azimuthal asymmetry which is due to multiple
interactions of partons at the initial stage of nuclear collisions,
and which is also present in $pA$ collisions. In our approach the
main source of azimuthal asymmetry is the combination of parton
multiple interactions with the steep variation of the nuclear
density at the edge of nuclei. We apply the light-cone dipole
formalism to compute the azimuthal asymmetry of prompt photons yield
from parton-nucleus, proton-nucleus and nucleus-nucleus collisions
at the RHIC energy.

\vspace{0.5cm}
\end{abstract}
\end{frontmatter}

\section{Introduction}
Prompt photons, i.e. photons not from hadronic decays, are interesting
since they do not participate in the strong interactions and therefore
carry information about the initial state hard
collisions. Nevertheless, measuring prompt photons is a challenge for
experimentalists, partly due to the existence of large backgrounds
coming from hadronic decays, which should be extracted.  Even after
this subtraction, there are other several sources for direct photons,
including thermal radiation from the hot medium and photons induced
by final state interactions with the medium. In this paper we
concentrate on the azimuthal asymmetry of prompt photons produced at
the initial stage of relativistic nuclear collisions.

The PHENIX collaboration at RHIC has recently reported in
Refs.~\cite{phe,phe2} the measurement of an
azimuthal asymmetry of direct photon production, which has been
studied recently in several theoretical papers
\cite{us-1,nd,nd-n,pd}. A novel mechanism which produces an
azimuthal asymmetry coming from the reaction's initial conditions
was introduced in Ref.~\cite{us-1}. This is in contrast with the
usual assumptions taken in approaches where the azimuthal asymmetry
is only associated with the properties of the medium created in the
final state. We show that at least part of the direct photon
azimuthal asymmetry, albeit small for $AA$ and $pA$ collisions,
originates from initial hard scatterings between partons of the
nuclei. In our approach, the main source of the azimuthal asymmetry
originates from the sensitivity of parton multiple interactions to
the steep variation of the nuclear density at the edge of the
nuclei, which correlates with the color dipole orientation.

This paper is organized as follows: In sections 2 and 3 we
introduce the main formalism for prompt photon production and
discuss the relevance of color dipole orientation. In section 4, we
introduce the azimuthal asymmetry for various collisions. In section
5, we present the numerical results. Some concluding remarks are
given in section 6.

\section{Photon radiation in the color dipole formalism }
The transverse momentum ($p_{T}$) distribution of photon
bremsstrahlung, coming from the interaction of a quark with nuclear
matter of thickness $T_{A}(b)=\int_{-\infty}^{\infty}dz
\rho_{A}(z,b)$ (where the nuclear density $\rho_{A}$ is integrated
along the parton trajectory at impact parameter $b$), integrated
over the final quark transverse momentum can be written as
\cite{us-1,kst1},
\begin{eqnarray}
\frac{d \sigma^{qA}(q\to q\gamma)}{d(ln \alpha)d^{2}\vec{p}_{T}d^{2}\vec{b}}&=&\frac{1}{(2\pi)^{2}}
\sum_{in,f}
\int d^{2}\vec{r}_{1}d^{2}\vec{r}_{2}e^{i \vec{p}_{T}.(\vec{r}_{1}-\vec{r}_{2})}
\phi^{\star}_{\gamma q}(\alpha, \vec{r}_{1})
\phi_{\gamma q}(\alpha, \vec{r}_{2})\nonumber\\
&\times&F_{A}(\vec{b},\alpha\vec{r}_{1},\alpha\vec{r}_{2},x),\label{m1}\
\end{eqnarray}
where $\alpha$ denotes the fraction of the quark light-cone momentum
carried by the photon, and $\phi_{\gamma q}(\alpha, \vec{r})$ is the
light-cone amplitude for the $q\gamma$ fluctuation with transverse
separation $\vec{r}$.  In this equation the QCD part  is encoded in
the function $F_{A}(\vec{b},\alpha\vec{r}_{1},\alpha\vec{r}_{2},x)$,
which is a linear combination of $\bar{q}q$ dipole partial
amplitudes on a nucleus at impact parameter $\vec{b}$,
\begin{eqnarray}
 F_{A}(\vec{b},\alpha\vec{r}_{1},\alpha\vec{r}_{2},x)&=&
\text{Im}f^{A}_{q\bar{q}}(\vec{b},\alpha\vec{r}_{1},x)+\text{Im}f^{A}_{q\bar{q}}(\vec{b},
\alpha\vec{r}_{2},x)\nonumber\\
&-&\text{Im}f^{A}_{q\bar{q}}(\vec{b},\alpha(\vec{r}_{1}-\vec{r}_{2}),x),\label{di}\
\end{eqnarray}
where the partial elastic amplitude $f^{A}_{q\bar{q}}$ can be
written, in the eikonal form, in terms of the dipole
elastic amplitude $f^{N}_{q\bar{q}}$ of a $\bar{q}q$ dipole
colliding with a proton at impact parameter $\vec{b}$ \cite{bk},
\begin{eqnarray}
\text{Im}f^{A}_{q\bar{q}}(b,\vec{r})&=&1-\Big[1-\frac{1}{A}\int d^{2}\vec{s}~ \text{Im}f^{N}_{q\bar{q}}
(\vec{s},\vec{r})T_{A}(\vec{b}+
\vec{s})\Big]^{A}\nonumber\\
&\approx&1-\exp[-\int d^{2}\vec{s}~ \text{Im}f^{N}_{q\bar{q}}(\vec{s},\vec{r})T_{A}(\vec{b}+\vec{s})].
\label{eik}\nonumber\\
\end{eqnarray}
Dependence on the light-cone momentum fraction $x$ of the target
gluons and $\alpha$ are implicit in the above expression. The dipole partial
elastic amplitude $f^{N}_{q\bar{q}}$ was proposed in
Ref.~\cite{us-1} to have the form
\begin{eqnarray}
\text{Im}f^{N}_{q\bar{q}}(\vec{s},\vec{r})&=&\frac{1}{12\pi}\int \frac{d^{2}\vec{q}}{q^{2}}
\frac{d^{2}\vec{q}^{\,\prime}}{q^{\prime 2}}
 e^{i\vec{s}.(\vec{q}-\vec{q}^{\,\prime})}\alpha_{s} \mathcal{F}(x,\vec{q},\vec{q}^{\,\prime})
\left(e^{-i\vec q\cdot\vec r\eta}-e^{i\vec q\cdot\vec r(1-\eta)}\right)\, \nonumber\\
&\times&\left(e^{i\vec q'\cdot\vec r\eta}-e^{-i\vec q'\cdot\vec r(1-\eta)}\right),
 \label{dung}\ 
\end{eqnarray} 
where we defined
$\alpha_{s}=\sqrt{\alpha_{s}(q^{2})\alpha_{s}(q^{\prime 2})}$.  The
fractional light-cone momenta of the quark and antiquark are denoted
by $\eta$ and $1-\eta$, respectively.  The radiated photon takes away
fraction $\alpha$ of the quark momentum. Therefore, we have the
parameter $\eta=1/(2-\alpha)$ for the photon production. It is known
that the center of gravity of $q\bar{q}$ is closer to fastest $q$ or
$\bar{q}$.  The generalized unintegrated gluon density\footnote{This
should not be mixed up with the generalized gluon density \cite{ji}
which is off-diagonal in the longitudinal fractional momentum $x$.}
$\mathcal{F}(x,\vec{q},\vec{q}^{\,\prime})$ is related to the diagonal
one by
$\mathcal{F}(x,q)=\mathcal{F}(x,\vec{q},\vec{q}=\vec{q}^{\,\prime})$.
Integrating over the vector $\vec{s}$ one can recover the dipole cross
section $\sigma^{N}_{q\bar{q}}(r)$, and also $\eta$ or $\alpha$
dependence will disappear
\begin{eqnarray}
\sigma^{N}_{q\bar{q}}(r)&=&2\int d^{2}\vec{s}~\text{Im}f^{N}_{q\bar{q}}(\vec{s},\vec{r})\nonumber\\
&=&\frac{4\pi}{3}\int\frac{d^{2}q}{q^{4}}(1-e^{-i\vec{q}.\vec{r}})\alpha_{s}(q^{2})\mathcal{F}(x,q).
\label{di-app}\
\end{eqnarray}

Relying on the saturation shape of the dipole cross-section $\sigma^{N}_{q\bar{q}}(r)$ \cite{kst},
the following form for
$\alpha_{s}\mathcal{F}(x,\vec{q},\vec{q}^{\,\prime})$ was proposed in Ref.~\cite{us-1},
\begin{eqnarray}
\alpha_{s}\mathcal{F}(x,\vec{q},\vec{q}^{\,\prime})&=&\frac{3\sigma_{0}}{16\pi^{2}}q^{2}
q^{\prime 2}R^{2}_{0}(x)e^{-\frac{1}{8}R^{2}_{0}(x)\left(q^{2}+q^{\prime 2}\right)}
e^{-\frac{1}{2}R^{2}_{N}\left(\vec{q}-\vec{q}^{\,\prime}\right)^{2}},\ \label{ung}
\end{eqnarray}
where the parameters $\sigma_{0}=23.03$ mb,
$R_{0}(x)=0.4\text{fm}\times (x/x_{0})^{0.144}$ with
$x_{0}=3.04\times 10^{-4}$ are fixed to HERA data for the proton
structure function \cite{kst}. The parameter
$R_{N}^{2}=5~\text{GeV}^{-2}$ is the $t$-slope of the pomeron-proton
vertex \cite{us-1}.  The energy scale $x$ which enters in the dipole
amplitude is related to the measurable variable $x=p_{T}/w$
\cite{amir2}, where $w$ is the center of mass energy. Unfortunately,
it is not possible to uniquely determine the unintegrated gluon
density function from the available data. Nevertheless, the proposed
form Eq.~(\ref{ung}) seems to be a natural generalization which
preserves the saturation properties of the diagonal part
\cite{us-1}. After carrying out the integrations in Eq.~(\ref{dung})
the dipole amplitude gets the following simple form,
\begin{eqnarray}
\text{Im}f^{N}_{q\bar{q}}(\vec{s},\vec{r})&=&\frac{\sigma_{0}}{8\pi B_{el}}
\Biggl\{\exp\left[-\frac{[\vec s+\vec r(1-\eta)]^2}{2B_{el}}\right] +
\exp\left[-\frac{(\vec s-\vec r\eta)^2}{2B_{el}}\right]\nonumber\\
&-&2\exp\Biggl[-\frac{r^2}{R_0^2(x)}-
\frac{[\vec s+(1/2-\eta)\vec r]^2}{2B_{el}}\Biggr]
\Biggr\},  \label{new-di}\
\end{eqnarray}
where we defined $B_{el}=R_{N}^{2}+R_{0}^{2}(x)/8$.

In Eq.~(\ref{m1}), $\phi_{\gamma q}(\alpha, \vec{r})$ is the
light-cone (LC) distribution amplitude of the projectile quark $\gamma
q$ fluctuation. Averaging over the initial quark polarizations and
summing over all final polarization states of the quark and photon, we get
\begin{eqnarray}
\sum_{in,f}\phi^{\star}_{\gamma q}(\alpha, \vec{r}_{1})\phi_{\gamma q}(\alpha, \vec{r}_{2})
&=& \frac{\alpha_{em}}{2\pi^{2}}m^2_{q}\alpha^{2}
\{\alpha^{2}K_{0}(\alpha m_{q} r_{1})K_{0}(\alpha m_{q} r_{2})\nonumber\\
&+&[1+(1-\alpha)^{2}]\frac{\vec{r}_{1}.\vec{r}_{2}}{r_{1}r_{2}}K_{1}( \alpha m_{q}r_{1})K_{1}
(\alpha m_{q} r_{2})\},\label{wave}\
\end{eqnarray}
where $K_{0,1}(x)$ denotes a modified Bessel function of the second
kind and $m_{q}$ is an effective quark mass, which can be conceived as
a cutoff regularization.  We take $m_{q}=0.2$ GeV for the case of
direct photon production \cite{amir2}. It has been also shown that a
value of $m_{q}=0.2$ GeV is needed in order to describe nuclear
shadowing effects \cite{mqq}.

Expression (\ref{m1}), with the exponentials expanded to first order
in the nuclear thickness, provides also the cross-section for direct
photon production in hadron-hadron collision. We have recently shown
that in this framework one can obtain a good description of the cross
section for prompt photon production data for proton-proton (pp)
collisions at RHIC and Tevatron energies
\cite{amir2}. Predictions for the LHC in the same framework are
given in Ref.~\cite{us-2}, while to compare with the predictions of
other approaches at the LHC see Ref.~\cite{lhc-hic}.

\section{Azimuthal asymmetry and dipole orientation}
\begin{figure}[!t]
               \centerline{\includegraphics[width=14.0 cm] {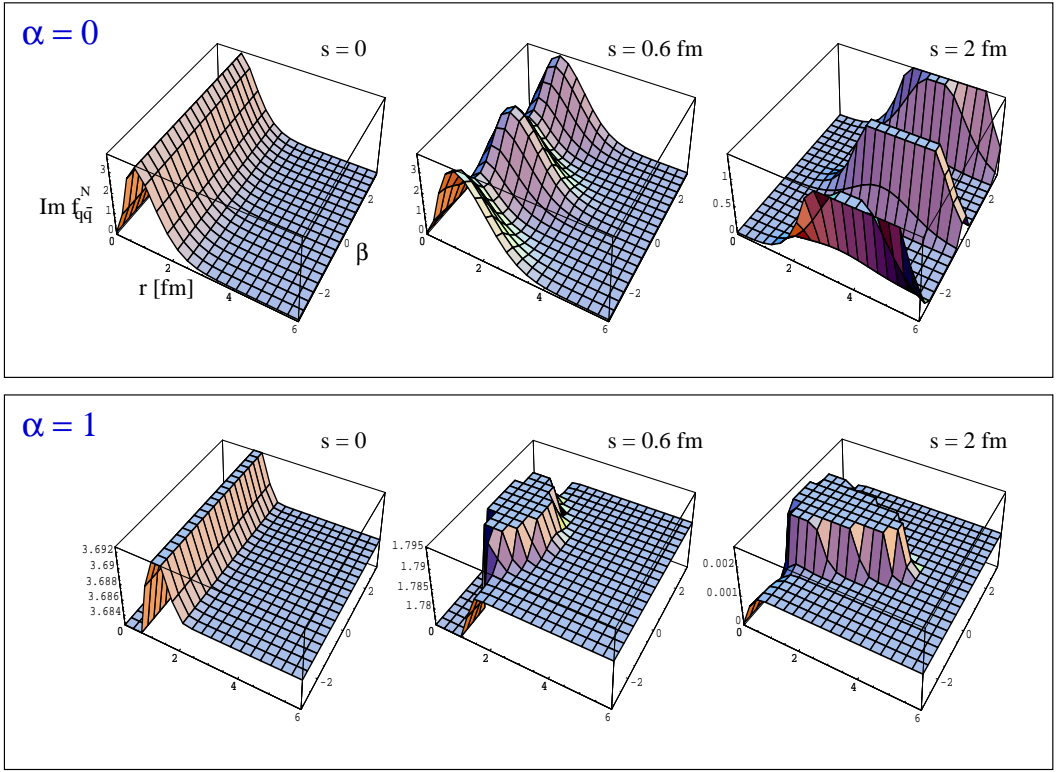}}
       \caption{The partial elastic amplitude
       $\text{Im}f^{N}_{q\bar{q}}$ (mb) of the $\bar{q}q$ dipole on a
       proton at impact parameter $s$ as a function of dipole size $r$
       and angle $\beta$ defined between $\vec{s}$ and $\vec{r}$ for two values of $\alpha=0, 1$. We
       use a fixed value of $x=0.01$ for all
       plots. \label{fig-4}}
\end{figure}
The main source of azimuthal asymmetry in the amplitude (\ref{eik})
is the interplay between multiple rescatering and the shape of the
physical system.  The key function which describes the effect of
multiple interactions is the eikonal exponential in Eq.~(\ref{eik}),
while the information about the shape of the system is incorporated
through a convolution of the impact parameter dependent partial
elastic amplitude and the nuclear thickness function. Notice that
the initial space-time asymmetry gets translated into a momentum
space anisotropy by the double Fourier transform in Eq.~(\ref{m1}).

It is quite obvious intuitively that although the Rutherford
scattering cross section is azimuthally symmetric, the azimuthal
angle of the radiated photons transverse momentum at a given impact
parameter $\vec{s}$ correlates with the direction of $\vec{s}$. In
terms of the partial elastic amplitude
$f^{N}_{q\bar{q}}(\vec{s},\vec{r})$, it means that the vectors
$\vec{r}$ and $\vec{s}$ are correlated. This is the key observation
which leads to an azimuthal asymmetry in pA and AA collisions. 

In Fig.~(\ref{fig-4}) we show the partial dipole amplitude
$f^{N}_{q\bar{q}}(\vec{s},\vec{r})$ as a function of the dipole size
$r$ and the angle $\beta$ between $\vec{s}$ and $\vec{r}$, at various
fixed values of $s$ for two values of $\alpha=0, 1$.
Notice that the generic feature of the partial dipole amplitude,
e. g. its maximum and minimum pattern, changes with $\alpha$. One can
see that for very small dipole size $r$ the dipole orientation is not
important. For very large dipole size $r$ compared to the impact
parameter $s$ or very small values of $s$ the dipole orientation is
also not present.  Notice, however, that it is not obvious a priori
how the convolution between the partial dipole amplitude and the
nuclear profile, which leads to a more complicated angle mixing, gives
rise to a final azimuthal asymmetry. The main aim of this paper is to
calculate such azimuthal asymmetry without using any approximations.

\section{Azimuthal asymmetry in $qA$, $pA$ and $AA$ collisions}
The azimuthal asymmetry of prompt photon production, resulting
from parton-nucleus (qA) collisions, is defined as the second order
Fourier coefficients in a Fourier expansion of the azimuthal
dependence of a single-particle spectra Eq.~({\ref{m1}) around the
beam direction,
\begin{equation}
v_{2}^{qA}(p_{T},b,
\alpha)= \frac{\int_{-\pi}^{\pi} d\phi \cos(2\phi)\frac{d
\sigma^{qA}(q\to q\gamma)}{d(ln \alpha)d^{2}\vec{p}_{T}d^{2}\vec{b}}}
{\int_{-\pi}^{\pi} d\phi \frac{d \sigma^{qA}(q\to q\gamma)}{d(ln
\alpha)d^{2}\vec{p}_{T}d^{2}\vec{b}}}, \label{v2-1}
\end{equation}
where the angle $\phi$ is defined with respect to the reaction
plane. Some of integrals in the above expression can be analytically
performed. After some tedious but straightforward calculation one
obtains
\begin{eqnarray}
&&v_{2}^{qA}(p_{T},b,\alpha)=
\frac{\int_{0}^{\infty} dr r\Psi_{N}(p_{T},r,\alpha)\Phi_{N}(b,r,\alpha)}
{\int_{0}^{\infty} dr r\Psi_{D}(p_{T},r,\alpha)\Phi_{D}(b,r,\alpha)+2\pi \mathcal{N}
(\alpha,p_{T})},\ \label{v2-form}
\end{eqnarray}
where the functions $\Psi_{D}$ and $\Psi_{N}$, which contain
information about $\gamma q$ fluctuation, are defined by
\begin{eqnarray}
\Psi_{N}(p_{T},r,\alpha)&=&\frac{\alpha_{em}}{2\pi^{2}}\{2m^{2}_{q}\alpha^{4}
\left(-\frac{J_{2}(p_{t}r)K_{0}(\alpha m_{q} r)}{p^{2}_{T}+(\alpha m_{q})^{2}}
+\frac{r}{4\alpha m_{q}} J_{2}(p_{t}r)K_{1}(\alpha m_{q} r)\right)\nonumber\\
&+&[1+(1-\alpha)^{2}]\Big(\frac{\alpha m_{q} p_{T}}{p_{T}^{2}+(\alpha m_{q})^{2}}
\left(J_{1}(p_{t}r)-J_{3}(p_{t}r)\right)K_{1}(\alpha m_{q} r)\nonumber\\
&+&J_{2}(p_{t}r)K_{0}(\alpha m_{q} r)-\frac{r\alpha m_{q}}{2}J_{2}(p_{t}r)K_{1}(\alpha m_{q} r)
\Big)\};\nonumber\\
\Psi_{D}(p_{T},r,\alpha)&=&\frac{\alpha_{em}}{2\pi^{2}}\{2m^{2}_{q}\alpha^{4}
\left(-\frac{J_{0}(p_{t}r)K_{0}(\alpha m_{q} r)}{p^{2}_{T}+(\alpha m_{q})^{2}}+\frac{r}{4\epsilon}
J_{0}(p_{t}r)K_{1}(\alpha m_{q} r)\right)\nonumber\\
&+&[1+(1-\alpha)^{2}]\Big(-\frac{2\alpha m_{q} p_{T}}{p_{T}^{2}+(\alpha m_{q})^{2}}J_{1}
(p_{t}r)K_{1}(\alpha m_{q} r)\nonumber\\
&+&J_{0}(p_{t}r)K_{0}(\alpha m_{q} r)-\frac{r\alpha m_{q} }{2}J_{0}(p_{t}r)K_{1}(\alpha m_{q} r)
\Big)\};\nonumber\\
\mathcal{N}(\alpha,p_{T})&=&\frac{\alpha_{em}}{2\pi^{2}}\left(\frac{m^{2}_{q}\alpha^{4}}{(p_{T}^{2}+
(\alpha m_{q})^{2})^{2}}
+\frac{(1+(1-\alpha)^{2})p_{T}^{2}}{(p_{T}^{2}+(\alpha m_{q})^{2})^{2}}\right),\nonumber\\  \label{di-psi}
\end{eqnarray}
where $J_{n}(x)~, n=0-3$ denotes the Bessel functions of the first
kind. The functions $\Phi_{N,D}(b,r,\alpha)$ in Eq.~(\ref{v2-form}) contain
information about the QCD part and also the shape of the nucleus;
\begin{eqnarray}
\Phi_{N}(b,r,\alpha)&=&-\int_{-\pi}^{\pi} d\beta e^{-\int d^{2}\vec{s}~\text{Im}f^{N}_{q\bar{q}}
(\vec{s},\alpha\vec{r})T_{A}(\vec{b}+\vec{s})} \cos(2\beta), \nonumber\\
\Phi_{D}(b,r,\alpha)&=&\int_{-\pi}^{\pi} d\beta e^{-\int d^{2}\vec{s}~\text{Im}f^{N}_{q\bar{q}}
(\vec{s},\alpha\vec{r})T_{A}(\vec{b}+\vec{s})},\ \label{ori}
\end{eqnarray}
where $\beta$ is the angle between $\vec{b}$ and $\vec{r}$, and the
dipole amplitude $f^{N}_{q\bar{q}}(\vec{s},\vec{r},)$ was defined in
Eq.~(\ref{new-di}). It is interesting to note that in the final form
of $v_{2}^{qA}$, Eq.~(\ref{v2-form}), the angle $\phi$ between the
impact parameter $\vec{b}$ and the transverse momentum of the
projectile quark $\vec{p}_{T}$ disappeared, and then the azimuthal
asymmetry is directly related to the dipole orientation with respect
to the impact parameter $\vec{b}$ through the angle $\beta$ (see
Eq.~(\ref{ori})). Therefore, if one neglects the dipole orientation
the azimuthal asymmetry becomes identically zero, regardless of both
the nuclear profile and the dipole cross-section parametrization.

In order to obtain the hadronic cross section from the elementary
partonic cross section Eq.~(\ref{m1}), we use the standard convolution
based on QCD factorization \cite{poorman},
\begin{eqnarray} &&\frac{d
\sigma^{\gamma}(pA\to \gamma X)}{dx_{F}d^{2}\vec{p}_{T}d^{2}\vec{b}}=
\frac{1}{x_{1}+x_{2}}\int_{x_{1}}^{1}\frac{d\alpha}{\alpha}
F_{2}^{p}(\frac{x_{1}}{\alpha},p_{T})
\frac{d \sigma^{qA}(q\to q\gamma)}{d(ln
\alpha)d^{2}\vec{p}_{T}d^{2}\vec{b}}\dot\ \label{con}
\end{eqnarray}
We take the parametrization for the proton structure function
$F_{2}^{p}(x,Q^{2})$ given in Ref.~\cite{ps}. Here $x_{1}$ denotes
the fraction of the light-cone momentum of the projectile hadron
carried away by the photon, and we define $x_{2}=x_{1}-x_{F}$, where
$x_{F}=2p_{L}/\sqrt{\bold{s}}$ is the Feynman variable. The
azimuthal asymmetry of photon yield in proton-nucleus ($pA$)
interactions is then defined as
\begin{equation} v_{2}^{pA}(b,p_{T})= \frac{\int_{-\pi}^{\pi}
d\phi \cos(2\phi) \frac{d \sigma^{\gamma}(pA\to \gamma
X)}{dx_{F}d^{2}\vec{p}_{T}d^{2}\vec{b}} } {\int_{-\pi}^{\pi}
d\phi\frac{d \sigma^{\gamma}(pA\to \gamma
X)}{dx_{F}d^{2}\vec{p}_{T}d^{2}\vec{b}} }, \label{v2-2}
\end{equation}
The above equation can be also simplified to an expression similar to
Eq.~(\ref{v2-form}), but now augmented with the proton structure function and an extra
integral over the variable $\alpha$.

The spectra of photon bremsstrahlung from nucleus-nucleus (AA)
collisions can be obtained from Eq.~(\ref{con}) by weighting the
cross-section with the density overlap factor of nuclei. In
principle, one may obtain the cross-section for AA collisions in the
same fashion as was done in Eq.~(\ref{con}) for the case of pA
collisions, that is by making a convolution with the nucleus
structure function instead of the proton structure function.
%
However, the medium modification of nucleon structure function for
the range of $p_{T}$ values we are interested in is less than $20\%$
\cite{e6} and therefore will not change the overall prediction. The
azimuthal asymmetry of photon yield from collisions of two nucleus
A$_{1}$ and A$_{2}$ at impact parameter $B$ is defined as
\begin{eqnarray}
v_{2}^{A_{1}A_{2}}(B,p_{T})&=&
\frac{\int_{-\pi}^{\pi} d\phi \cos(2\phi)~\mathcal{G}_{N}}
{\int_{-\pi}^{\pi} d\phi~\mathcal{G}_{D}};\nonumber\\
\mathcal{G}_{N}&=& \int d^{2}\vec{b}
\cos(2\Theta_{1}) \frac{d\sigma^{\gamma}(pA_{1}\to \gamma
X)}{dx_{F}d^{2}\vec{p}_{T}d^{2}\vec{b}_{1}} T_{A_{2}}(\vec{b}_{2})\nonumber\\
&+& \int d^{2}\vec{b}\cos(2\Theta_{2}) \frac{d\sigma^{\gamma}(pA_{2}\to \gamma
X)}{dx_{F}d^{2}\vec{p}_{T}d^{2}\vec{b}_{2}} T_{A_{1}}(\vec{b}_{1});
\nonumber\\
\mathcal{G}_{D}&=&\int d^{2}\vec{b} \left(\frac{d \sigma^{\gamma}(pA_{1}\to \gamma
X)}{dx_{F}d^{2}\vec{p}_{T}d^{2}\vec{b}_{1}} T_{A_{2}}(\vec{b}_{2})
+\frac{d \sigma^{\gamma}(pA_{2}\to \gamma
X)}{dx_{F}d^{2}\vec{p}_{T}d^{2}\vec{b}_{2}} T_{A_{1}}(\vec{b}_{1})\right),
\label{v2-3}\nonumber\\
\end{eqnarray}
 where we used the notation $\vec{b}_{2}=\vec{b}+\vec{B}$,
 $\vec{b}_{1}=\vec{b}$ ($\vec{b}$ is the impact parameter of the p$A_{1}$
 collision) and the angle $\Theta_{1}$ ($\Theta_{2}$) is the
 angle between the vectors $\vec{b}_{1}$( $\vec{b}_{2}$) and $\vec{B}$,
 respectively. The factor $\cos(2\Theta_{1,2})$ in the above equation
 relates the reaction planes of the pA and the AA collisions.  The
 integral over $\vec{b}$ in Eq.~(\ref{v2-3}) covers the almond shape
 area of the nucleus-nucleus overlap.

\section{Numerical result and discussions}
We will perform a numerical calculation for the RHIC energy
$\sqrt{\bold{s}}=200$ GeV, at midrapidities. The only external input
is the nuclear profile. First, we take a popular Woods-Saxon (WS)
profile, with a nuclear radius $R_{A}=6.5$ fm and a surface
thickness $\xi=0.54$ fm, for Pb+Pb collisions \cite{ws}.
\begin{figure}[!t]
       \centerline{\includegraphics[width=9 cm] {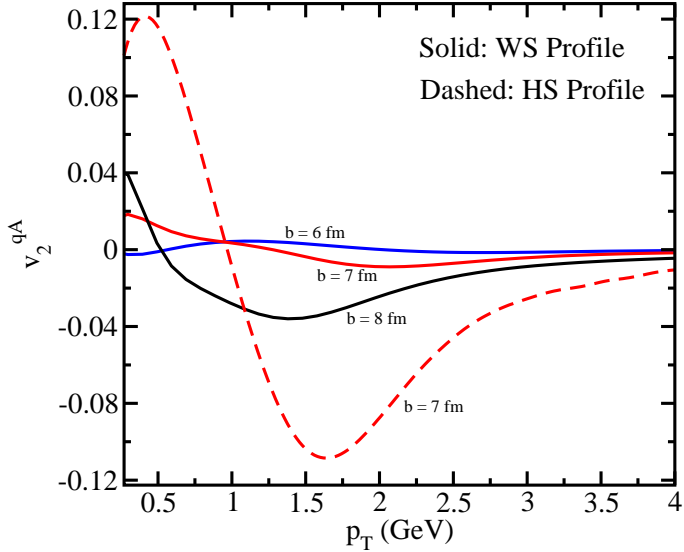}}
       \caption{
       The azimuthal anisotropy of prompt photon production
       coming from quark-nucleus collisions,
       at various impact parameter $b$ and at the RHIC energy,  for both the Woods-Saxon (WS)
       and hard sphere (HS) profiles. The relative fraction of
       the quark momentum carried by photon is taken to be $\alpha=1$ for all curves. \label{fig-1}}
\end{figure}

In Fig.~(\ref{fig-1}) we show the calculated values of $v_{2}^{qA}$
defined in Eq.~(\ref{v2-1}), for fixed $\alpha=1$,  at various qA
collision impact parameters $b$, and at the RHIC energy.  For
central collisions, the correlation between nuclear profile and
dipole orientation is minimal. In fact, if the nuclear profile
function was constant, then the convolution between the nuclear
profile and the dipole orientation, defined in Eq.~(\ref{eik}),
would be trivial, and $v_{2}^{qA}$ becomes then identically zero.
Therefore, the main source of azimuthal anisotropy is not present
for central collisions. This can be seen in Fig.~(\ref{fig-1}),
where a pronounced elliptic anisotropy is observed for collisions
with impact parameters close to the nuclear radius $R_{A}$, where
the nuclear profile undergoes rapid changes. Therefore, an important
parameter which controls the elliptic asymmetry in this mechanism is
$|b-R_{A}|$. We have verified this numerically by taking different
$R_{A}$ values for the WS profile, but with the same surface
thickness $\xi$.

 \begin{figure}[!t] \centerline{\includegraphics[width=9 cm]
 {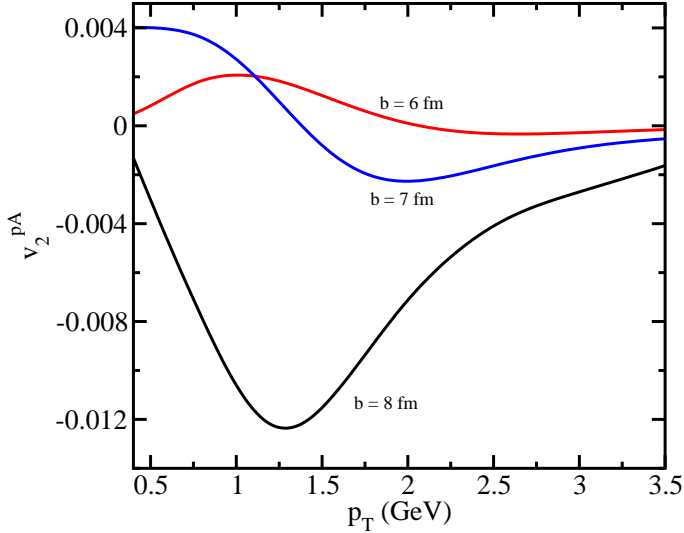}} \caption{The impact parameter dependence of prompt
 photon azimuthal asymmetry for proton-nucleus collisions at various
 impact parameter $b$ at RHIC energy. For the nuclear profile, we have
 taken the Woods-Saxon (WS) profile for all curves.  \label{fig-2}}
\end{figure}

In Fig.~(\ref{fig-2}) we show $v_{2}^{pA}$ for prompt photons
produced in non-central pA collisions, for the WS nuclear profile.
Notice that at very small transverse momentum our results are not
reliable, because the employed proton structure function is not valid.

In Fig.~(\ref{fig-3}) we show the prompt photon azimuthal asymmetry for AA
collisions defined in Eq.~(\ref{v2-3}), at various impact parameters and at
RHIC energy.
The absolute value of $v_{2}^{AA}$ turns out to be reduced compared
to both $v_{2}^{pA}$ and $v_{2}^{qA}$. The reason is that the
integrand in Eq.~(\ref{v2-3}) gets contributions only from
semi-peripheral pA collisions where our mechanism is at work, and
most of the integral over $\vec{b}$ does not contribute. This
significantly dilutes the signal. For prompt photon, there is no
suppression mechanism related to medium effects, as for the case of
hadron production in central AA collisions compared to pp collisions
\cite{star-s}.
For hadronic $v_{2}$, the inclusion of such suppression effects
significantly enhances the azimuthal asymmetry coming from this
mechanism \cite{amir1}. The other diluting factor for $v_{2}^{AA}$
is the presence of an extra $\cos(2 \Theta)$ in Eq.~(\ref{v2-3}),
which accounts for the changing of the reaction plane going from pA
to AA collisions. At high $p_{T}$, where the dipole size is very
small, the dipole orientation becomes less important and
consequently the correlation between the dipole cross-section and
the nuclear profile disappears, i.e. the azimuthal asymmetry
vanishes. This can be seen from
Figs.~(\ref{fig-1},\ref{fig-2},\ref{fig-3}), where $v_{2}$ for all
qA, pA and AA collisions approaches to zero at high $p_{T}$.

 \begin{figure}[!t]
       \centerline{\includegraphics[width=9 cm] {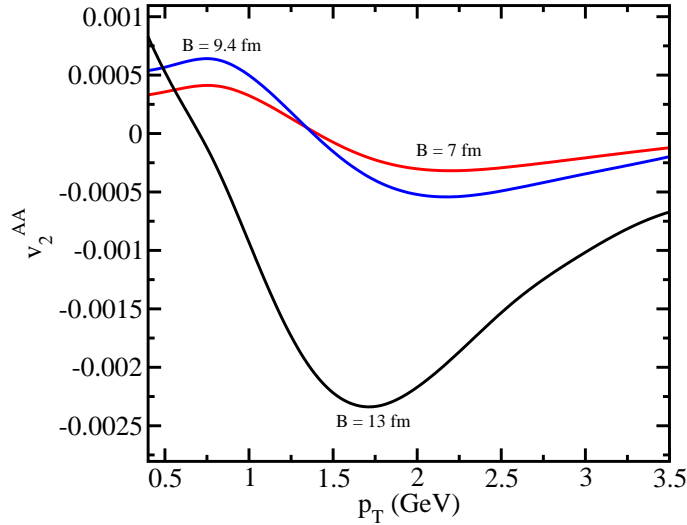}}
       \caption{The impact parameter $B$ dependence of prompt photon
       azimuthal asymmetry, for Pb+Pb collisions at RHIC energy, with
       the WS nuclear profile.\label{fig-3}}
\end{figure}

In our approach the profile of nuclear density at the edge is a very
important input, since the elliptic asymmetry stems from the rapid
change of nuclear density at the edge.  In order to show this more
clearly and to estimate the theoretical uncertainty of our
calculations, we obtain the elliptic asymmetry for the hard sphere
profile with a constant density distribution,
$\rho_{A}=\rho_{0}\Theta(R_{A}-r)$, with the same nuclear radius
$R_{A}$ as the WS profile for Pb.  In Fig.~(\ref{fig-1}) we show the
azimuthal asymmetry $v_{2}^{qA}$ for the hard sphere profile (HS) at
impact parameter $b=7$ fm. For the HS profile the nuclear thickness
changes at nuclear radius more steeply compared to the WS profile
and consequently the elliptic asymmetry is significantly bigger.
Notice that at impact parameter $b=7$ fm for the HS profile, even
though there is no matter, the correlation between color dipole and
the variation of nuclear thickness still exists and consequently the
azimuthal asymmetry is not zero.

\section{Summary and final remarks}
We have computed the azimuthal asymmetry of prompt photons originating
from primary hard scatterings between partons. This can be accounted
for by the inclusion of the color dipole orientation, which is
sensitive to the rapid variation of the nuclear profile. We showed
that the azimuthal asymmetry $v_2$ coming from this mechanism changes
the sign and becomes negative for peripheral collisions, albeit it is extremely small.
The first experimental attempts by PHENIX to extract the elliptic
flow of direct photons yielded results which are compatible with
zero within error bars \cite{phe}. However, the data was still
contaminated with background from hadron decays. Recently, the
PHENIX collaboration has presented preliminary data on $v_{2}$ of
direct photons \cite{phe2}. Although the systematics errors are
still very large, the data indicates that $v_{2}$ of direct photons
is larger than our prediction for $v_{2}$ of prompt photons.
Therefore, if data is confirmed, other sources for $v_{2}$ of direct
photons should be also important.


There are, however, a number of caveats in our approach which need
further study before taking the numbers predicted here at face
value. As we have shown, the tail of nuclear profile is an important
external input in this mechanism and quite significantly affects the
results. For example, we showed that the azimuthal asymmetry is
enhanced almost by an order of magnitude for the HS nuclear profile
compared to the WS one (see Fig.~(\ref{fig-1})). Unfortunately the
tail of all available nuclear profile parametrizations is less
reliable, since it is not well probed by electron scattering and is
obtained by a simple extrapolation. This is also due to the fact
that the neutron distribution, which may be more important on the
periphery, cannot be properly accounted for by electron scattering
data. This brings uncertainty to our results. Furthermore, as we
already pointed out, due to lack of experimental data, there is some
freedom left to define the off-diagonal part of the unintegrated
gluon density. Although it is not important for the total
cross-section, it plays an important role for the azimuthal
asymmetry.

\section*{Acknowledgments}
We are grateful to Hans-J\"urgen Pirner for many illuminating
discussions. AHR would like also to thank Francois Gelis, Jamal
Jalilian-Marian and Kirill Tuchin for useful discussions. This work
was supported in part by Fondecyt (Chile) grants 1070517 and 1050589
and by DFG (Germany) grant PI182/3-1.


\begin{thebibliography}{20}

\bibitem{phe}
PHENIX Collaboration, arXiv:0705.1711.
\bibitem{phe2}
PHENIX Collaboration, Phys. Rev .Lett. {\bf 96}, 032302 (2006).
\bibitem{us-1}
 B. Z. Kopeliovich, H. J. Pirner, A. H. Rezaeian, Ivan Schmidt, Phys. Rev. {\bf D77}, 034011 (2008)[arXiv:0711.3010].
\bibitem{nd}
 S. Turbide, C. Gale and R. J. Fries, Phys. Rev. Lett. {\bf 96}, 032303 (2006).
\bibitem{nd-n}
 S. Turbide, C. Gale, E. Frodermann, U. Heinz, Phys. Rev. {\bf C77}, 024909 (2008) [arXiv:0712.0732].
\bibitem{pd}
 R. Chatterjee, E. S. Frodermann, U. W. Heinz, D. K. Srivastava, Phys .Rev. Lett. {\bf 96}, 202302 (2006).
\bibitem{kst1}
B.Z. Kopeliovich, A. Schaefer and A.V. Tarasov,
Phys. Rev. {\bf C59} (1999) 1609
\bibitem{bk}
B.~Z.~Kopeliovich,
Phys.\ Rev.\  C {\bf 68}, 044906 (2003).
\bibitem{ji}
X. Ji, Phys. Rev. Lett. {\bf 78}, 610 (1997); Phys. Rev. {\bf D55}, 7114 (1997).
A. V. Radyushkin, Phys. Lett. {\bf B380}, 417 (1996); Phys. Rev. {\bf D56}, 5524 (1997).
\bibitem{kst}
K. Golec-Biernat and M. W\"usthoff, Phys. Rev. {\bf D59}, 014017 (1999).
\bibitem{amir2}
 B. Z. Kopeliovich, A. H. Rezaeian, H. J. Pirner and Ivan Schmidt, Phys. Lett. {\bf B653}, 210 (2007).
\bibitem{mqq}
B. Z. Kopeliovich, J. Raufeisen and A. V. Tarasov, Phys. Rev. {\bf C62}, 035204 (2000).
\bibitem{us-2}
A. H. Rezaeian, B. Z. Kopeliovich, H. J. Pirner and Ivan Schmidt, arXiv:0707.2040.
\bibitem{lhc-hic}
 S. Abreu {\em et al.}, arXiv:0711.0974.
\bibitem{poorman}
 B. Z. Kopeliovich, J. Raufeisen, A. V. Tarasov, Phys. Lett. {\bf B503}, 91 (2001).
\bibitem{ps}
SMC Collaboration, Phys. Rev. {\bf D58}, 112001 (1998).
\bibitem{ws}
A. Bohr and B. R. Mottelson, {\it Nuclear Structure} (Benjamin, New York, 1969);
C. W. De Jager, H. De Vries, C. De Vries, Atom. Data Nucl. Data Tables {\bf 36}, 496 (1987).
\bibitem{e6}
E665 Collaboration, Z. Phys. {\bf C67}, 403 (1995).
\bibitem{star-s}
STAR Collaboration, Phys. Rev. Lett. {\bf 89}, 202301 (2002).
\bibitem{amir1}
 B. Z. Kopeliovich, A. H. Rezaeian, and Ivan Schmidt, under preparation.


\end{thebibliography}
\end{document}